\documentclass[10pt,conference]{IEEEtran}
\usepackage[T1]{fontenc}
\usepackage[hidelinks]{hyperref}
\usepackage{color}

\usepackage{cite}
\usepackage{amsmath,amssymb,amsfonts}
\usepackage{graphicx}
\usepackage{textcomp}
\usepackage{xcolor}
\usepackage{balance}
\def\BibTeX{{\rm B\kern-.05em{\sc i\kern-.025em b}\kern-.08em
    T\kern-.1667em\lower.7ex\hbox{E}\kern-.125emX}}

\usepackage{pifont}
\newcommand{\cmark}{\ding{51}}%
\newcommand{\xmark}{\ding{55}}%

\usepackage{booktabs}
\usepackage[caption=false, font=footnotesize]{subfig}

\usepackage{orcidlink}
\usepackage{url}
\usepackage[inline]{enumitem}
\usepackage{svg}

\newcommand{\requirement}[1]{\textbf{R#1}}

\begin{document}
\bstctlcite{MyBSTcontrol}
\title{Bridging HCI and AI Research for the Evaluation of Conversational SE Assistants}
\author{\IEEEauthorblockN{Jonan Richards \orcidlink{0009-0007-1218-8599}, Mairieli Wessel \orcidlink{0000-0001-8619-726X}}
\IEEEauthorblockA{\textit{Radboud University} \\
Nijmegen, The Netherlands\\
jonan.richards@ru.nl, mairieli.wessel@ru.nl}
}

\maketitle

\begin{abstract}
As Large Language Models (LLMs) are increasingly adopted in software engineering, recently in the form of conversational assistants, ensuring these technologies align with developers' needs is essential. The limitations of traditional human-centered methods for evaluating LLM-based tools at scale raise the need for automatic evaluation. In this paper, we advocate combining insights from human-computer interaction (HCI) and artificial intelligence (AI) research to enable human-centered automatic evaluation of LLM-based conversational SE assistants. We identify requirements for such evaluation and challenges down the road, working towards a framework that ensures these assistants are designed and deployed in line with user needs.
\end{abstract}


\section{Introduction}

A variety of bots have been used in software engineering (SE) to automate tasks and increase developers' productivity \cite{erlenhov2019CurrentFutureBots}. Large Language Models (LLMs) show much promise in powering these automations, increasingly supporting various activities, including code generation, code summarization, bug detection, and requirements analysis \cite{hou2024LargeLanguageModels}. Recent research has demonstrated the potential of integrating multiple tasks in a single SE bot to reduce developers' cognitive overload \cite{wessel2022BotsPullRequests}. A similar trend is seen in the emergence of LLM-based ``assistants'' capable of supporting several SE activities simultaneously, both in industry \cite{li2024SoftwareEngineeringFoundation,pinto2024DeveloperExperiencesContextualized} and research \cite{ross2023ProgrammersAssistantConversational}.

Implementing such a multi-faceted assistant as a conversational agent is a logical choice \cite{ross2023ProgrammersAssistantConversational}. LLM-based conversational assistants increase productivity over non-conversational tools by allowing developers to work towards their goals incrementally \cite{austin2021ProgramSynthesisLarge}. Moreover, they can facilitate developers' learning \cite{choudhuri2024HowFarAre}, and natural language provides an intuitive mechanism for users to give feedback \cite{erlenhov2019CurrentFutureBots}. Developers greatly value conversational interaction in LLM-based SE assistants \cite{pinto2024DeveloperExperiencesContextualized,ross2023ProgrammersAssistantConversational}.

As the capabilities of bots in SE evolve, constant evaluation is needed to ensure they are designed and deployed in accordance with users' needs, which tools for SE often fail to do \cite{lo2015HowPractitionersPerceive}. Human-centered methods from the field of human-computer interaction (HCI) include any method that involves humans, and are necessary for this alignment with users' needs \cite{myers2016ProgrammersAreUsers}. An important aspect of using human-centered methods is to design in iterations, while evaluating the tool with experts or users \cite{myers2016ProgrammersAreUsers}. Evaluation with users is the most direct evidence of the performance of an SE tool \cite{davis2023WhatNotWorking} and is the gold standard for conversational agents \cite{allouch2021ConversationalAgentsGoals}. However, user studies are costly, time-intensive, and difficult to scale \cite{davis2023WhatNotWorking}.

LLM-based SE tools, and specifically conversational assistants, suffer from a lack of robust evaluation methods \cite{hou2024LargeLanguageModels,pinto2024DeveloperExperiencesContextualized}. LLMs are highly sensitive to the phrasing of their prompts, or instructions, and the design of an LLM-based tool takes much ``prompt engineering'' to achieve the desired functionality \cite{pan2024HumanCenteredDesignRecommendations}. As LLMs are non-deterministic and allow unrestricted textual interaction, human-centered methods such as expert analyses of LLMs' responses \cite{jiang2022PromptMakerPromptbasedPrototyping} and user studies are infeasible for the scale and frequency of evaluation that is required during the design process. Since LLM-based tools for SE are commonly evaluated on benchmark datasets \cite{hou2024LargeLanguageModels}, a quick and more systematic alternative is evaluating small changes in prompts on a subset of a benchmark. However, LLMs' responses to benchmark inputs are traditionally compared to corresponding reference outputs using similarity metrics such as BLEU, ROUGE, and METEOR \cite{hou2024LargeLanguageModels}. These reference-based metrics fail to capture how LLM-based tools perform in practice and do not correspond well to human evaluation of natural language \cite{liu2016HowNOTEvaluate}.

The high cost of user studies has led to the emergence of an alternative evaluation method in the field of HCI. LLM-based simulated users have been used to qualitatively evaluate interfaces \cite{xiang2024SimUserGeneratingUsability}, conversational agents \cite{dewit2024LeveragingLargeLanguage}, and also have the potential to be used for automatic, synthesized user studies of SE tools \cite{gerosa2024CanAIServe}. The limitations of traditional benchmarks for evaluating natural language generation and the infeasibility of performing user studies to guide prompt engineering have led to a similar development in the field of artificial intelligence (AI). In LLM-as-a-Judge approaches \cite{zheng2024JudgingLLMjudgeMTbench}, an LLM is employed as a ``judge'', replacing real humans to assess LLM-generated texts automatically. LLM-as-a-Judge has increasingly been used by practitioners to evaluate domain-specific LLM-based tools \cite{li2024SoftwareEngineeringFoundation}, and may be helpful in various SE tasks as well \cite{ahmed2024CanLLMsReplace}. Although both discussed methods leverage LLMs to synthesize research data, they reflect the fields from which they have emerged. Simulated users have been employed in HCI research to generate artificial interaction data and qualitative insights, whereas LLM-as-a-Judge approaches are used as quantitative benchmarks in AI research. While simulated users have the potential to produce quantitative metrics and LLM-as-a-Judge approaches may offer qualitative insights, these areas are under-explored.

In this paper, we advocate combining these two methods from different fields to enable human-centered automatic evaluation of conversational LLM-based assistants for SE. We identify several requirements for such evaluation, discuss LLM-as-a-Judge approaches and simulated users, and propose combining insights from both methods into an approach we envision will address all requirements. Finally, we identify challenges to address when implementing the proposed approach. These findings were derived from an extensive literature search. Our contribution is these insights towards realizing scalable and comprehensive automatic evaluation of conversational LLM-based developer assistance, meant to not replace but complement manual evaluation in the human-centered design process.

\section{Requirements}
In this section, we describe several requirements for the automatic evaluation of conversational SE assistants.

\renewcommand\thesubsectiondis{R\arabic{subsection}.}
\subsection{Explore realistic conversations}
\begin{figure}[htbp]
    \centering
    \includegraphics[width=\columnwidth]{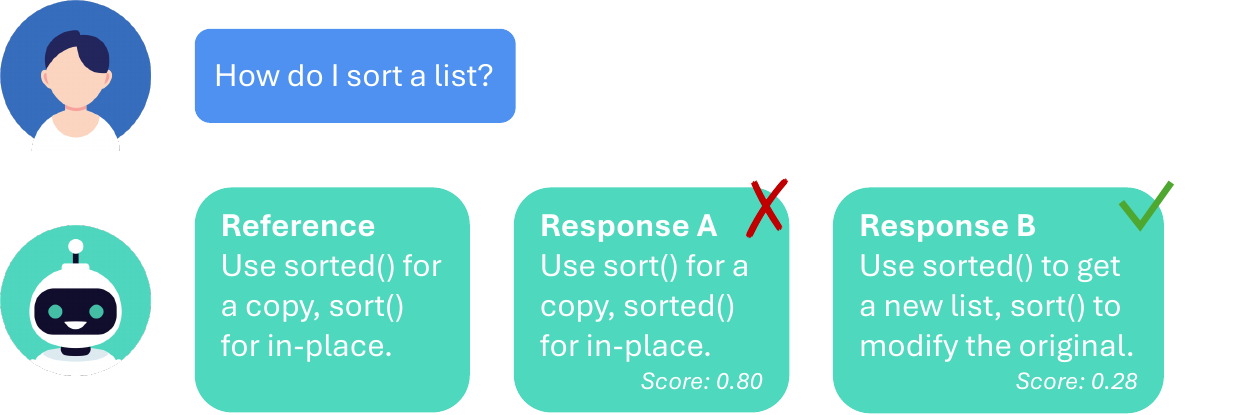}
    \caption{Flaws in reference-based evaluation: incorrect response A scores high on BLEU-4, while correct response B scores low due to phrasing.}
    \label{fig:reference}
    \vspace{-.5\baselineskip}
\end{figure}

Typically, benchmarks for large-scale evaluation of conversational agents come in either of two forms: single-turn question and answer pairs, or multi-turn dialogues \cite{allouch2021ConversationalAgentsGoals}. However, single-turn interactions do not capture conversational dynamics spanning multiple turns, leaving a large part of the interaction space unexplored. Both multi-turn and single-turn dialogue datasets typically provide reference responses, ignoring the wide range of responses that may be valid and the many ways in which a conversation might unfold \cite{allouch2021ConversationalAgentsGoals}. Additionally, evaluation of generated text based on similarity to a reference does not correspond well to how humans judge natural language \cite{liu2016HowNOTEvaluate}. \autoref{fig:reference} demonstrates how reference-based evaluation of SE assistants' responses may lead to flawed insights. Thorough evaluation of conversational SE assistants, therefore, requires reference-free exploration of realistic interactions over multiple turns.

\subsection{Support diversity}
One individual's needs may not fully correspond to another's. Problem-solving tools often best support the information processing styles and motivation of young, male users, while other users' perspectives are neglected
\cite{burnett2016GenderMagMethodEvaluating}. This can lead to the occurrence of ``inclusivity bugs'', which are issues with the usability of a tool that are triggered only for specific groups of users \cite{guizani2022HowDebugInclusivity}. LLMs are especially vulnerable to these inclusivity bugs, as their inherent unpredictability may cause disruptions for one user but not for others \cite{amershi2019}. Instances of inclusivity bugs have been encountered in user studies of conversational LLM-based SE assistants, being triggered based on experience level \cite{nam2024}, gender \cite{choudhuri2024HowFarAre}, or interaction style \cite{richards2024WhatYouNeed}. Prompt engineering can be used to put ``guardrails'' on interactions between LLMs and users to remove biases and ensure fairness \cite{dong2024SafeguardingLargeLanguage}. First, however, conversational SE assistants need to be evaluated from the perspective of a diverse, representative user base to reliably detect inclusivity bugs.

\subsection{Provide quantitative metrics}
Incremental changes to LLMs' prompts and parameters are enabled by constant evaluation and comparison \cite{pan2024HumanCenteredDesignRecommendations}. Quantitative measures are invaluable for this benchmarking process, allowing for quick assessment of whether a change improved or reduced a tool's performance \cite{jiang2022PromptMakerPromptbasedPrototyping}. However, the sensitivity of LLMs to its input may cause evaluation to reflect superficial aspects of the prompt's phrasing \cite{thomas2024LargeLanguageModels}. Evaluating a large number of paraphrased prompts instead would allow designers to effectively assess the impact of changing the underlying structure of a prompt \cite{thomas2024LargeLanguageModels,jiang2022PromptMakerPromptbasedPrototyping}, making the prototyping process of an LLM-based SE assistant more systematic and robust. Applying such techniques, however, requires automatic quantitative metrics. Finally, quantitative benchmarks are useful in demonstrating the effectiveness of new assistants \cite{myers2016ProgrammersAreUsers} and in comparing them with existing solutions.

\subsection{Provide qualitative insights}
While quantitative metrics do allow a designer to quickly assess the impact of a change, they do not offer any information on usability issues or potential improvements. Conversational agents are often tested with users to gather qualitative feedback throughout the design process \cite{sadek2023TrendsChallengesProcesses}. LLM-based SE assistants are no different, and feedback is essential to ensure these assistants are developed in line with users' needs \cite{pinto2024DeveloperExperiencesContextualized}. Automatic generation of qualitative feedback on SE assistants would allow designers to anticipate and fix usability issues \cite{gerosa2024CanAIServe}, complementing user studies.

\renewcommand\thesubsectiondis{\Alph{subsection}.}

\section{Current situation}
Here, we discuss user simulations and LLM-as-a-Judge approaches and how these methods address the identified requirements. A comparison of the methods can be found in \autoref{table:comparison}.

\begin{table}[htbp]
    \vspace{-.5\baselineskip}
    \caption{Comparison between simulated users and LLM-as-a-Judge}
    \label{table:comparison}
    \setlength\tabcolsep{0pt}
    \centering
    \begin{tabular*}{\columnwidth}{@{\extracolsep{\fill}}l c c}
        \toprule
        Requirement & Simulated Users & LLM-as-a-Judge \\
        \midrule
        R1. Explore realistic conversations & \cmark & \xmark \\
        R2. Support diversity & \cmark & \cmark \\
        R3. Provide quantitative metrics & \textbf{?} & \cmark \ \\
        R4. Provide qualitative insights & \cmark & \textbf{?} \\
        \bottomrule
    \end{tabular*}
    \vspace{-.5\baselineskip}
\end{table}

\subsection{Simulated Users}
Recently, researchers have explored leveraging LLMs to simulate humans and generate synthetic research data. Hämäläinen et al. \cite{hamalainen2023EvaluatingLargeLanguage} found that GPT-3 was able to produce realistic qualitative data, substantially overlapping with data generated by humans and even producing additional insights. LLM-simulated simulated user interactions were found to be realistic, and useful for assessing the effects of design choices during prototyping of social media platforms \cite{tornberg2023SimulatingSocialMedia}. Xiang et al. \cite{xiang2024SimUserGeneratingUsability} used an LLM to simulate a user interacting with an interface, and found it was useful for identifying edge-cases, generating a substantial amount of usability feedback not identified by human users. These studies demonstrate simulated users can be leveraged for qualitative insights (\requirement{4}), both by allowing designers to inspect interactions and by directly generating feedback.

An application of user simulations with especially much potential is in the evaluation of conversational agents. Simulated users do not rely on static benchmark datasets but can be interactively prompted, and can produce realistic, believable dialogue \cite{dewit2024LeveragingLargeLanguage,jin2024TeachTuneReviewingPedagogical} (\requirement{1}). De Wit \cite{dewit2024LeveragingLargeLanguage} found that researchers were able to identify several usability issues in a conversational agent by leveraging simulated user interactions.

When instructed to impersonate a persona, or description of a user, simulated users can align their manner of interaction with the persona's characteristics \cite{xiang2024SimUserGeneratingUsability,jin2024TeachTuneReviewingPedagogical}. When employing simulated users with a broad range of personas, designers can support diversity by identifying inclusivity bugs \cite{tornberg2023SimulatingSocialMedia} (\requirement{2}).

There has been little research on producing quantitative insights through user simulations. De Wit \cite{dewit2024LeveragingLargeLanguage} proposes having simulated users generate synthetic survey data to measure user experience of a conversational agent. ChatGPT has been able to replicate outcomes of some quantitative SE surveys, but this has not been attempted for user studies of SE tools \cite{steinmacher2024CanChatGPTEmulate}. As of yet, it is unclear whether user simulations can provide reliable quantitative metrics, and if so, how (\requirement{3}).

\subsection{LLM-as-a-Judge}

Zheng et al. \cite{zheng2024JudgingLLMjudgeMTbench} coined the term ``LLM-as-a-Judge'', referring to approaches where LLMs are used as a ``judge'' to evaluate LLM-generated texts. These approaches come in several forms including grading a single LLM output, or selecting the best of a pair of outputs \cite{zheng2024JudgingLLMjudgeMTbench}. These judgments can be made based on a set of criteria \cite{pan2024HumanCenteredDesignRecommendations}, meaning LLM-as-a-Judge can be employed to provide a wide range of quantitative metrics (\requirement{3}). The evaluation provided by LLM-as-a-Judge approaches often align well with human judgments \cite{zheng2024JudgingLLMjudgeMTbench}, including for various SE tasks \cite{ahmed2024CanLLMsReplace}. In addition to their high scalability and low cost compared to human evaluation, this has led to LLM-as-a-Judge being increasingly employed in practice for evaluating LLM-based tools \cite{li2024SoftwareEngineeringFoundation}. 

Furthermore, judge LLMs can be instructed to provide an explanation for their assessment \cite{zheng2024JudgingLLMjudgeMTbench}. However, there is a lack of research on the usefulness of these explanations, and whether they can be used to pinpoint usability issues, provide feedback, and thus generate qualitative insights \requirement{4}).

\begin{figure}[htbp]
    \centering
    \subfloat[Reference-free dataset]{
        \includegraphics[width=0.45\linewidth]{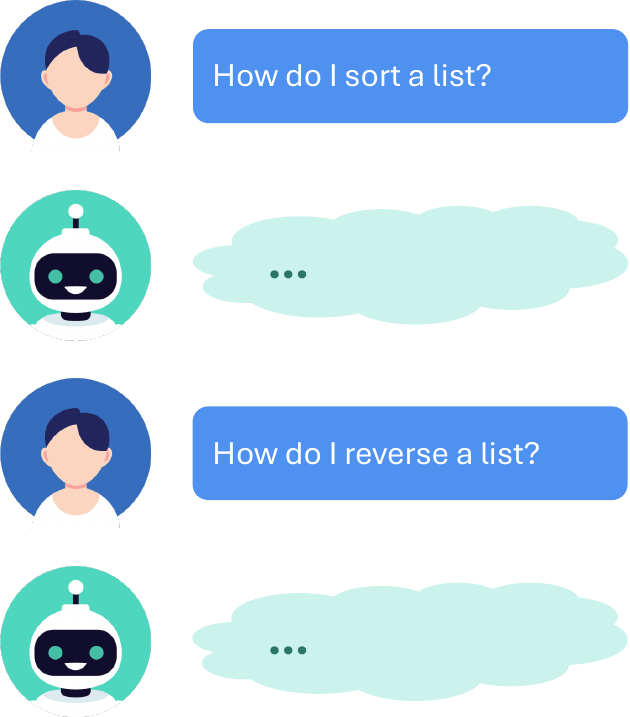}
    }
    \hfil
    \subfloat[Realistic interaction]{
        \includegraphics[width=0.45\linewidth]{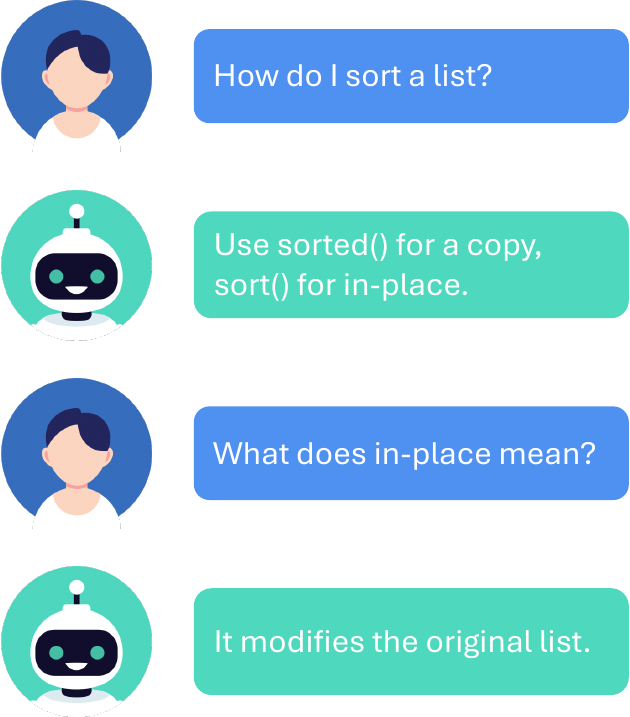}
    }
    
    \caption{Reference-free datasets miss conversations where responses refer to earlier messages.}
    \label{fig:conv}
    \vspace{-.5\baselineskip}
\end{figure}

Although LLM-as-a-Judge approaches can be used to evaluate outputs without a reference, datasets of inputs are still needed to serve as test cases. When evaluating conversational agents using multi-turn datasets, the lack of reference outputs makes it impossible to anticipate all potential interactions, as demonstrated in \autoref{fig:conv}. These reference-free datasets are likely not representative of real users' interactions (\requirement{1}), who might ask follow-up questions an SE assistants' earlier response. Besides, constructing domain- and tool-specific input datasets for LLM-as-a-Judge may cost substantial effort. When designing an LLM-based SE assistant, changing or adding features would involve updating this dataset to accommodate for new interactions, slowing down the prototyping process. 

Through fine-tuning on human feedback, LLMs have been aligned with the preferences of an ``average user'' \cite{wang2024ArithmeticControlLLMs}. While input datasets can be constructed to reflect the interactions of a diverse group of users, LLMs employed as judges will only evaluate responses from the point of view of this average user. However, LLMs can predict individuals' preferences regarding natural language when provided personal context in the form of personas \cite{hu2024QuantifyingPersonaEffect}. Providing personas is effective in LLM-as-a-Judge settings too, allowing judge LLMs to achieve performance comparable to human annotators in predicting individuals' assessments \cite{dong2024CanLLMBe}. Therefore, LLM-as-a-Judge approaches can support diversity when leveraging representative interaction datasets and user personas (\requirement{2}).

\section{Towards a solution}

\subsection{Proposed method}
We highlight that the two discussed methods complement each other well. Bridging the gaps between HCI and AI research by combining the two methods can fulfill the discussed requirements and provide automatic human-centered evaluation of LLM-based conversational assistants for SE. We are the first to propose combining simulated users and LLM-as-a-Judge. Although providing LLM-as-a-Judge with a persona could arguably be seen as analogous to having a simulated user give quantitative assessments, the novelty in the proposed method lies in its framing that enables the transfer of practical insights between the fields of HCI and AI.

Employing the proposed combined method, one should start by creating a representative set of personas and relevant evaluation criteria, both quantitative and qualitative. As designers using LLM-as-a-Judge approaches have been found to adapt and add evaluation criteria in response to new insights \cite{pan2024HumanCenteredDesignRecommendations}, we imagine the set of personas may evolve similarly throughout the design process. A round of evaluation using the proposed method would require the following steps, for each persona:

\begin{enumerate}[label=\arabic*)]
    \item Generate interactions between the assistant and a simulated user leveraging the persona.
    \item Generate qualitative feedback on the interaction with a simulated user leveraging the persona.
    \item Employ LLM-as-a-Judge for a quantitative assessment based on the persona and evaluation criteria.
\end{enumerate}

Note that we do not intend for this method to replace expert analysis or user studies. Rather, we expect it will complement the design process by enabling quick evaluation during prototyping and increasing the scope of evaluation with real users. Repeated comparison of the proposed method with pilot user studies is necessary to ensure that the personas, simulated interactions, and evaluation align with real users.

\subsection{Challenges}
\renewcommand\thesubsubsectiondis{C\arabic{subsubsection})}
We identify several challenges that must be addressed before successfully applying the proposed evaluation method to conversational SE assistants.

\subsubsection{Create representative personas}
Simulated users will need to be representative of the intended end users. Personas described in the literature may serve as a starting point for these simulations. The GenderMag method \cite{burnett2016GenderMagMethodEvaluating} describes personas based on several facets of using problem-solving tools, and a study at Microsoft \cite{ford2017CharacterizingSoftwareEngineering} characterized software engineers' personas based on differences in tasks, collaboration styles, and perspectives on autonomy. More research into individuals' intents, interaction styles, and preferences regarding conversational SE assistants is needed to extend and adapt these personas to the context of conversational SE assistants, increasing the realism of simulated users' interactions and evaluations through LLM-as-a-Judge. 

\subsubsection{Identify and mitigate biases}
LLMs have been found to produce biased and stereotyped outputs when simulating humans \cite{argyle2023OutOneMany,taubenfeld2024SystematicBiasesLLM} or when used in LLM-as-a-Judge approaches \cite{zheng2024JudgingLLMjudgeMTbench}. There are techniques to mitigate such biases \cite{argyle2023OutOneMany,wang2024LargeLanguageModels,zheng2024JudgingLLMjudgeMTbench}, and biased user simulations may still prove useful in increasing the scale and coverage of evaluation as long as they do not replace real user studies \cite{wang2024LargeLanguageModels}. However, the proposed method requires thorough testing to mitigate potential biases and ensure that its effects are not harmful.

\subsubsection{Handle context}
There is much potential in conversational SE assistants providing tailored support by leveraging local context such as code repositories, documentation, and information about the current task \cite{melo2023SupportingContextualConversational,pinto2024DeveloperExperiencesContextualized}. Techniques such as Retrieval-Augmented Generation (RAG) have extensively been researched to provide relevant context to LLMs while handling limitations on input length \cite{gao2024RetrievalAugmentedGenerationLarge}. Although there is a wide range of metrics to evaluate RAG \cite{gao2024RetrievalAugmentedGenerationLarge}, more research is needed on how these can be integrated with user simulations and LLM-as-a-Judge approaches.

\section{Concluding discussion}
Besides the discussed challenges, more research is needed regarding implementation choices such as prompt design, temperature settings, and the number and length of conversations. However, we expect existing research on simulated users \cite{xiang2024SimUserGeneratingUsability,jin2024TeachTuneReviewingPedagogical} and LLM-as-a-Judge \cite{zheng2024JudgingLLMjudgeMTbench} to provide useful guidance on these implementation choices.

Although employing an LLM for evaluation of another LLM may seem counterintuitive, existing literature provides evidence for the effectiveness of this approach \cite{zheng2024JudgingLLMjudgeMTbench}. However, the performance of LLM-as-a-Judge approaches varies based on the SE task being evaluated \cite{ahmed2024CanLLMsReplace}, and simulated users occasionally suffer from hallucination \cite{xiang2024SimUserGeneratingUsability} and may not fully capture the cognitive processes of behavior and cognition \cite{dewit2024LeveragingLargeLanguage}. To assess the viability of the combined method proposed in this paper and explore the potential of LLM-based evaluation in SE, we intend to employ it during the design of an LLM-based conversational SE assistant, and compare the synthesized results to those of a user study.

In should be noted that the goal is not to replace the human aspect by relying only on automatic methods for evaluation. Instead, the proposed approach is intended to complement manual human-centered methods by extending the scale, coverage, and frequency of evaluation. By allowing for comparison and alignment with human judgments, we consider this method to be an automatic human-centered approach. Still, it must be extensively tested and employed carefully to uphold ethical considerations and human-centered priorities.

To conclude, we have identified several requirements for automatic human-centered evaluation of LLM-based conversational assistants for SE and discuss whether simulated users and LLM-as-a-Judge address these requirements. We proposed a method combining both approaches and identified challenges down the road. In this way, we contribute to aligning future SE assistants with developers' needs.

\balance
\bibliographystyle{IEEEtran}
\bibliography{patch,references}
\end{document}